\documentclass[]{pasj01}

\usepackage{natbib}
\usepackage{time}

\Received{2017/8/12}
\Accepted{yyyy/mm/dd}

\begin{document} 

\title{A Challenge to Identify an Optical Counterpart of the Gravitational Wave Event GW151226 with Hyper Suprime-Cam\thanks{Based on data collected at Subaru Telescope, which is operated by the National Astronomical Observatory of Japan.}}

\author{Yousuke \textsc{Utsumi}\altaffilmark{1}}
\altaffiltext{1}{Hiroshima Astrophysical Science Center, Hiroshima University, 1-3-1 Kagamiyama, Higashi-Hiroshima, Hiroshima, 739-8526, Japan}
\email{youtsumi@hiroshima-u.ac.jp}

\author{Nozomu \textsc{Tominaga}\altaffilmark{2,3}}
\altaffiltext{2}{Department of Physics, Faculty of Science and Engineering, Konan University, 8-9-1 Okamoto, Kobe, Hyogo 658-8501, Japan}
\altaffiltext{3}{Kavli Institute for the Physics and Mathematics of the Universe (WPI), The University of Tokyo Institutes for Advanced Study, The University of Tokyo, 5-1-5 Kashiwa, Chiba 277-8583, Japan}

\author{Masaomi \textsc{Tanaka}\altaffilmark{4}}
\altaffiltext{4}{National Astronomical Observatory of Japan, 2-21-1 Osawa, Mitaka, Tokyo 181-8588, Japan}

\author{Tomoki \textsc{Morokuma}\altaffilmark{5}}
\altaffiltext{5}{Institute of Astronomy, Graduate School of Science, The University of Tokyo, 2-21-1 Osawa, Mitaka, Tokyo 181-0015, Japan}

\author{Michitoshi \textsc{Yoshida}\altaffilmark{6}}
\altaffiltext{6}{Subaru Telescope, National Astronomical Observatory of Japan, 650 North A'ohoku Place, Hilo, HI 96720, USA}

\author{Yuichiro \textsc{Asakura}\altaffilmark{7,$\dagger$}}
\altaffiltext{7}{Institute for Space-Earth Environmental Research, Nagoya University, Furo-cho, Chikusa-ku, Nagoya, Aichi 464-8601, Japan}
\altaffiltext{$\dagger$}{Deceased 18 August 2017}

\author{Fran\c{c}ois \textsc{Finet}\altaffilmark{6}}

\author{Hisanori \textsc{Furusawa}\altaffilmark{4}}

\author{Koji S. \textsc{Kawabata}\altaffilmark{1}}

\author{Wei \textsc{Liu}\altaffilmark{1,8}}
\altaffiltext{8}{University of Chinese Academy of Sciences, No.2 Beijing West Road, Purple Mountain Observatory, Nanjing, 210008, China}

\author{Kazuya \textsc{Matsubayashi}\altaffilmark{9}}
\altaffiltext{9}{Okayama Astrophysical Observatory, National Astronomical Observatory of Japan, 3037-5 Honjou, Kamogata, Asakuchi, Okayama 719-0232, Japan}

\author{Yuki \textsc{Moritani}\altaffilmark{3}}

\author{Kentaro \textsc{Motohara}\altaffilmark{5}}

\author{Fumiaki \textsc{Nakata}\altaffilmark{6}}

\author{Kouji \textsc{Ohta}\altaffilmark{10}}
\altaffiltext{10}{Department of Astronomy, Kyoto University, Kitashirakawa-Oiwake-cho, Sakyo-ku, Kyoto, Kyoto 606-8502, Japan}

\author{Tsuyoshi \textsc{Terai}\altaffilmark{6}}

\author{Makoto \textsc{Uemura}\altaffilmark{1}}

\author{Naoki \textsc{Yasuda}\altaffilmark{3}}

\author{on behalf of the J-GEM collaboration}


\maketitle

\begin{abstract}
We present the results of the detailed analysis of an optical imaging survey conducted using the Subaru / Hyper Suprime-Cam (HSC),
which aims to identify an optical counterpart to the gravitational wave event GW151226.
In half a night, the $i$- and $z$-band imaging survey by HSC covers 63.5deg$^2$ of the error region,
which contains about 7\% of the LIGO localization probability, and the same field is observed in three different epochs.
The detectable magnitude of the candidates in a differenced image is evaluated as $i \sim 23.2$ mag for the requirement of at least two 5$\sigma$ detections,
and 1744 candidates are discovered.
Assuming a kilonova as an optical counterpart, we compared the optical properties of the candidates with model predictions.
A red and rapidly declining light curve condition enables the discrimination of a kilonova from other transients,
and a small number of candidates satisfy this condition.
The presence of stellar-like counterparts in the reference frame suggests that the surviving candidates are likely to be flare stars.
The fact that most of those candidates are in galactic plane, $|b|<5^{\circ}$, supports this interpretation.
We also checked whether the candidates are associated with the nearby GLADE galaxies,
which reduces the number of contaminants even with a looser color cut.
When a better probability map (with localization accuracy of $\sim50{\rm deg}^2$) is available, 
kilonova searches of up to approximately $200$ Mpc will become feasible by conducting immediate follow-up observations with an interval of 3--6 days.
\end{abstract}

\newpage

\section{Introduction}
Following the first detection of the gravitational waves (GWs) from a  36 and 29$M_{\odot}$ binary black hole coalescence
by the LIGO interferometer \citep[GW150914:][]{2016PhRvL.116f1102A}, 
a series of studies on binary black hole coalescences, GW151226 \citep{2016PhRvL.116x1103A} and GW170104 \citep{2017arXiv170601812T} were subsequently reported.
An increasing number of GW events is expected because efforts on improving the performance of LIGO are still underway,
and large efforts on the construction and commissioning of Advanced Virgo \citep{2015CQGra..32b4001A} and KAGRA \citep{2012CQGra..29l4007S}
are ongoing. Also, the LIGO-India project \citep{2013IJMPD..2241010U} has been approved\footnote{http://www.gw-indigo.org/tiki-index.php}.
Although the estimates for compact-binary coalescence rates depend on a number of assumptions and uncertain model parameters, the likely binary neutron-star detection rate for the Advanced LIGO-Virgo network is 40 events per year, with a range between 0.4 and 400 per year \citep{2010CQGra..27q3001A}.

The search for electromagnetic (EM) counterparts to GW events has become the next milestone for GW astronomy.
EM follow-ups will help us understand the broad physical picture of the GW sources.
\citet{1998ApJ...507L..59L} first proposed that radioactive decay energy gives rise to emissions in the UV-optical-IR wavelength range in neutron-star mergers.
These emissions have been called ``Macronovae" \citep{2005astro.ph.10256K} or ``Kilonovae" \citep{2010MNRAS.406.2650M}.
Kilonova emissions are expected to have shorter durations and redder colors than typical supernovae (SNe)
because of the high opacities of the Lanthanide elements \citep{2013ApJ...775...18B,2013ApJ...774...25K,2013ApJ...775..113T}.
However, in reality, it is challenging to conduct follow-up optical searches for EM counterparts because of the low localization accuracy of the GW events.
For example, the areas of localization are approximately 230 deg$^2$ (90\% credible region) for GW150914
and  850 deg$^2$ for GW151226 \citep{2016PhRvX...6d1015A},
whereas the typical maximum field-of-view (FoV) of an optical search with a large telescope  is a few square deg.

Extensive EM follow-up observation campaigns in multiple wavelengths have been organized 
to cover the large location uncertainties \citep{2016ApJ...826L..13A,2016ApJS..225....8A}.
Once  LIGO detects a GW signal, a notice is immediately distributed to EM searchers 
via  a private network established on the Gamma-ray Coordinates Network\footnote{http://www.ligo.org/scientists/GWEMalerts.php} immediately.
The notice contains a probability sky map  and its false alarm rate, which is a measure of the confidence level.
With this information, EM searchers can conduct their observations.
We participate in this framework through the Japanese collaboration for Gravitational wave EM follow-ups (J-GEM),
utilizing the optical and radio telescopes in Japan, as well as those in New Zealand, China, South Africa, Chile, and Hawaii \citep{2016PASJ...68L...9M}.
Initiated by the notice for GW151226, J-GEM conducted coordinated EM searches
using the J-GEM facilities, which includes the Subaru / Hyper Suprime-Cam (HSC) \citep{2017PASJ...69....9Y}.
HSC is a very-wide-field imager installed on the prime focus of the 8.2m Subaru telescope atop Maunakea
\citep{2012SPIE.8446E..0ZM}.
Its FoV is a circular aperture with a diameter of 1.5$^{\circ}$, which is the largest among the currently existing 8-10 m telescopes,
providing the potential to detect magnitudes as faint as 26.
In this paper, we focus on the HSC observation for GW151226, especially the detection of faint candidates.
This is the first HSC follow-up observation for optical counterparts of the GW events.

There is little expectation of a detectable EM signature from GW151226, because it  has been revealed to be a binary black hole merger.
In fact, \citet{2017PASJ...69....9Y} concluded that no optical counterpart was detected with their HSC observation,
which was mostly based on visual inspection.
However, investigating the method of identifying an optical counterpart to a GW event from an HSC survey is important
for future follow-up observations, because we will face a large number of contaminations by non-GW transient sources.
A deep and wide-field transient survey will detect a large number of candidates at once.
\citet{2008ApJ...676..163M} first presented the results of a wide and deep transient survey in a 0.91 deg$^2$ field over the Subaru / XMM-Newton Deep Field \citep[SXDF;][]{2008ApJS..176....1F}
with five pointings with the Subaru / Suprime-Cam, whose total survey area is slightly smaller but comparable to one HSC pointing.
The detection limit for variable components was $i\sim25.5$ mag.
Their transients were classified based on the optical morphology, magnitude, color, host galaxy optical-to-mid-infrared color,
variable component spatial offset from the host galaxy, and light curve.
They detected 1040 transients including variable stars, SNe, and AGNs from their 8-10 epochs over 2-4 years of observation.
These results indicate that there are many possible
candidates that could contaminate the transient sample.

In contrast to previous surveys,  only a few number of visits is allowed in an HSC imaging survey for a GW counterpart because of the limited observing time.
We therefore investigate a way to identify an optical counterpart to a GW event with limited information.
In this paper, we outline our strategy and present the results of the analysis with actual data taken for GW151226.

\section{HSC observation}\label{selection}

\subsection{HSC observation for GW151226}
\begin{figure}
 \begin{center}
  \includegraphics[width=\linewidth]{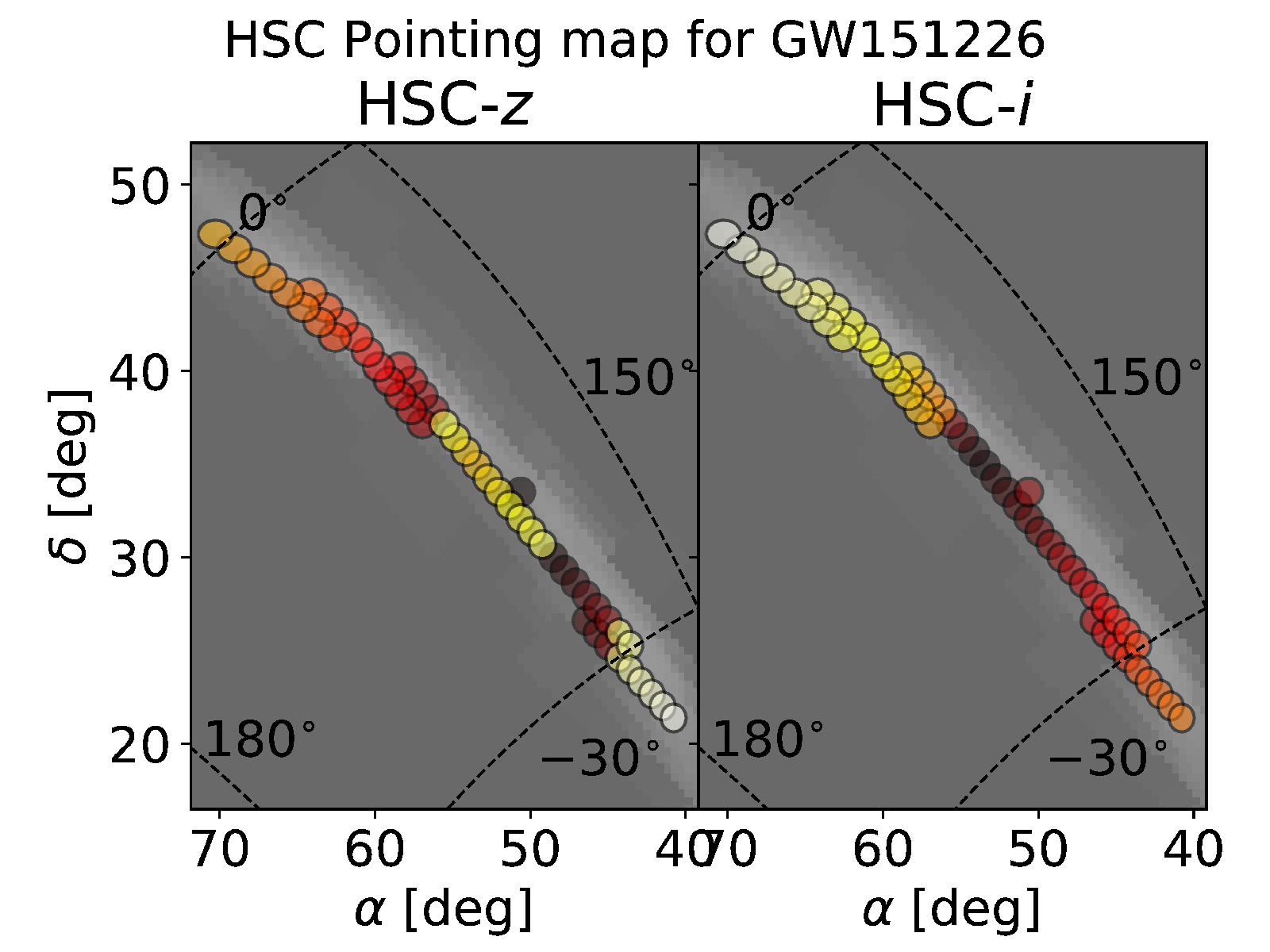} 
 \end{center}
\caption{Pointing map for GW151226 overlaid on the refined  probability map \citep[\emph{LALInference\_skymap.fits.gz};][]{GCN18889}.
The circles represent the FoV of the HSC, with the colors indicating the sequence of observations for the first epoch (Jan. 7, 2016).
Observations were made from the darkest color to the lightest color. Details of the observation process are described in appendix \ref{plan}.
The dashed curves represent the Galactic graticules.}\label{fig:pointingmap}
\end{figure}

The Target-of-Opportunity (ToO) observation of GW151226 \citep[2015-12-26 03:38:53.648 UTC;][]{GCN18728} with HSC was scheduled three times in the first half of the nights of  Jan. 7, 13, and Feb. 6, 2016 (UT).
We used only two filters to minimize the overhead introduced by the filter exchanging sequence.
The selected filters were the $i$- and $z$-bands, because the kilonova emission is expected to be red in color \citep{2013ApJ...775...18B,2013ApJ...774...25K,2013ApJ...775..113T}.
The observation started with an integration time of 60 s,
but was reduced to 45 s to make the survey area as wide as possible.
As it is impossible to cover the entire 1400 deg$^2$ of the 90\% credible region suggested by the LIGO probability map \citep{GCN18728},
we focused on a small patch of the most probable region.
We adopted the probability map \emph{bayestar.fits.gz} \citep{2016PhRvD..93b4013S}  distributed for GW151226.
Figure \ref{fig:pointingmap} shows the pointing layout overlaid on the probability map.
Fifty pointings are defined on the HEALPix  \citep{2005ApJ...622..759G} grid distributed over the most probable region 
allowing the FoV to overlap each other.
The adopted resolution of the HEALPix is NSIDE=64, which corresponds to 0.84 deg$^2$/pix$^2$.
We define these pointings by referring to the initial map, and the probability map is finally refined \citep{GCN18889}.
As a result, the pointings are revealed to deviate slightly from the ridge of the most probable region.
Each pointing is visited twice in an epoch with a slight offset of 1 arcmin to identify and remove cosmic rays and moving objects in both the $i$ and $z$ filters.
We describe the limitations of the HSC observation in appendix \ref{limitations}
and the details of the observations in \citet{2017PASJ...69....9Y} and in appendix \ref{plan}.

\subsection{Data reduction and quality assurance}
We ran \emph{hscPipe} v4.0.1 to reduce the images,
which is a standard reduction pipeline \citep{2017arXiv170506766B} developed for the HSC Strategic Survey Program \citep[HSC SSP;][]{2017arXiv170405858A}
available to general observers.
\emph{hscPipe} removes instrumental signatures by de-biasing,
cosmic ray removal, fixing CCD defects, applying cross talk correction,
correcting brighter-fatter effects,  flat fielding, illumination correction,
and astrometric and photometric calibrations.
Astrometric and photometric calibrations were performed against the PanSTAARS catalog \citep{2016arXiv161205560C}.
After removing the instrumental signatures of all the individual CCDs, \emph{hscPipe} generates the stacked images,
warping individual images with the resulting astrometric solution by
dividing the entire survey field with a collection of $1.7\times1.7~{\deg}^2$ subregions, which are called tracts.
A tract consisting of $9\times 9$ sub-subregions is called a ``patch''.
We call the stacked images, ``pre-differenced'' images.
We generated a pre-differenced image for  each patch, filter, and observing date.

\begin{table*}[htdp]
\tbl{Summary of the HSC observation.}{
\begin{tabular}{cccccccc}
\hline
Date (UT)           & Start-T$^{*}$     &   Filter  &   $m_{\rm lim, 5\sigma}$      &       &    Seeing(${}^{\prime\prime}$)    &       &     Note\\
                &   [days]          &           &            [AB]           &   min     &    average    &   max    &     \\
\hline
2016/01/07      &     12.06      &      HSC-$z$   &   $23.8\pm0.2$        &   0.42    &   0.60    &   0.91    & 1st     \\
2016/01/07      &     12.19      &      HSC-$i$   &   $24.6\pm0.2$        &   0.57    &   0.96    &   1.71    &         \\
2016/01/13      &     18.05      &      HSC-$z$   &   $23.8\pm0.2$        &   0.48    &   0.65    &   1.11    & 2nd     \\
2016/01/13      &     18.17      &      HSC-$i$   &   $24.6\pm0.2$        &   0.55    &   0.97    &   1.68    &         \\  
2016/02/06      &     42.06      &      HSC-$z$   &   $23.8\pm0.3$        &   0.42    &   0.60    &   0.84    & Reference    \\
2016/02/06      &     42.18      &      HSC-$i$   &   $24.4\pm0.2$        &   0.54    &   0.71    &   1.12    &         \\
\hline
\end{tabular}}\label{tab:stacked}
\begin{tabnote}
\footnotemark[$*$] Start-T indicates the number of days to start the observation since the LIGO trigger.
\end{tabnote}
\end{table*}%

We evaluated the data quality of the pre-differenced images for each observing date in terms 
of the seeing and limiting magnitude.
The seeing is derived from the pre-differenced images.
To quantify the depth of our data, we calculated the limiting magnitude as follows.
We randomly distributed 1.5-arcsec apertures on the sky area  avoiding source regions
and derived the standard deviation of the distribution of the derived values
by a given circular aperture photometry as an estimator of the sky fluctuation.
The size of the aperture is larger than the average seeing size of the Subaru / HSC by a factor of about two.
We set a $5\sigma$ value of the sky fluctuation as the limiting magnitude.
The derived values are listed in table \ref{tab:stacked}.
These limiting magnitudes are apparently deeper compared to the other optical deep and wide-field imaging surveys
for optical counterpart searches such as DECam \citep{2016ApJ...826L..29C} and PanSTARRS \citep{2016ApJ...827L..40S}.
We also derived the achieved survey area by counting if each HEALPix grid is inside our HSC pointings.
We represent the HSC FoV as a circular shape with a diameter of 1.5$^\circ$ for simplicity.
The adopted resolution of the HEALPix grid is NSIDE=512. The achieved survey area is 63.5 deg$^2$.
The probability that the location of the GW event is covered by this HSC observation is about 7\%.

\subsection{Image differencing}

\begin{table*}[htdp]
\tbl{ 5sigma limiting magnitude of difference images.}{
\begin{tabular}{ccc}
	\hline
	Operands				&	$i_{\rm lim,5\sigma,diff}$	&	$z_{\rm lim,5\sigma,diff}$	\\
	\hline
	1st-ref		&	$24.3\pm0.2$	&	$23.6\pm0.2$	\\
	2nd-ref		&	$24.3\pm0.2$	&	$23.5\pm0.2$	\\
	\hline
\end{tabular}}\label{tab:qualityfordifferencedimage}
\begin{tabnote}
\end{tabnote}
\end{table*}%

To identify transient candidates, we adopted the image differencing technique.
We subtractedd the reference images taken on Feb. 6 from the images taken on the 1st epoch (Jan. 7) and the 2nd epoch (Jan. 13) by homogenizing the size of the point spread function (PSF).
Then, we applied the same tools of the detection and measurement available
in \emph{hscPipe} to the four differenced images to identify transient candidates and obtain the four magnitudes: $i_{\rm 1st},i_{\rm 2nd},z_{\rm 1st},z_{\rm 2nd}$.
We also derived the 1.5-arcsec-aperture limiting magnitudes for the differenced images (table \ref{tab:qualityfordifferencedimage}).
 This is an approximate measure of magnitude of the faintest source detected.

We adopt the following criteria to identify a source as ``detected" on one of the four differenced images.
\begin{enumerate}
\item $|(S/N)_{\rm PSF}| > 5$ in the differenced image is required to  confirm a high confidence level.
\item elongation / (elongation)$_{\rm PSF}$ $> 0.8$ is required to detect a stellar-like source.
\item $0.8 < {\rm FWHM}/{\rm (FWHM)_{PSF}} < 1.3$ is required to also detect a stellar-like source.
\item PSF-subtracted residual $< 3\sigma$ (in a differenced image) is required to confirm that the source can be described by a PSF.
\item $S/N({\rm 1.5^{\prime\prime}}) > 5$ is not required in a differenced image; however, it is required in the pre-differenced image to confirm a detection in the pre-differenced image.
\end{enumerate}
We define $N_{\rm det}$  for each source by counting the number of detections if the source satisfies the above criteria for different epochs and filter bands. $N_{\rm det}=2$ if the source is detected in both the filters in a night, or if the source is detected in both epochs but not in specific filter bands.

\begin{figure}
 \begin{center}
  \includegraphics[width=8cm,bb=0 0 461 345]{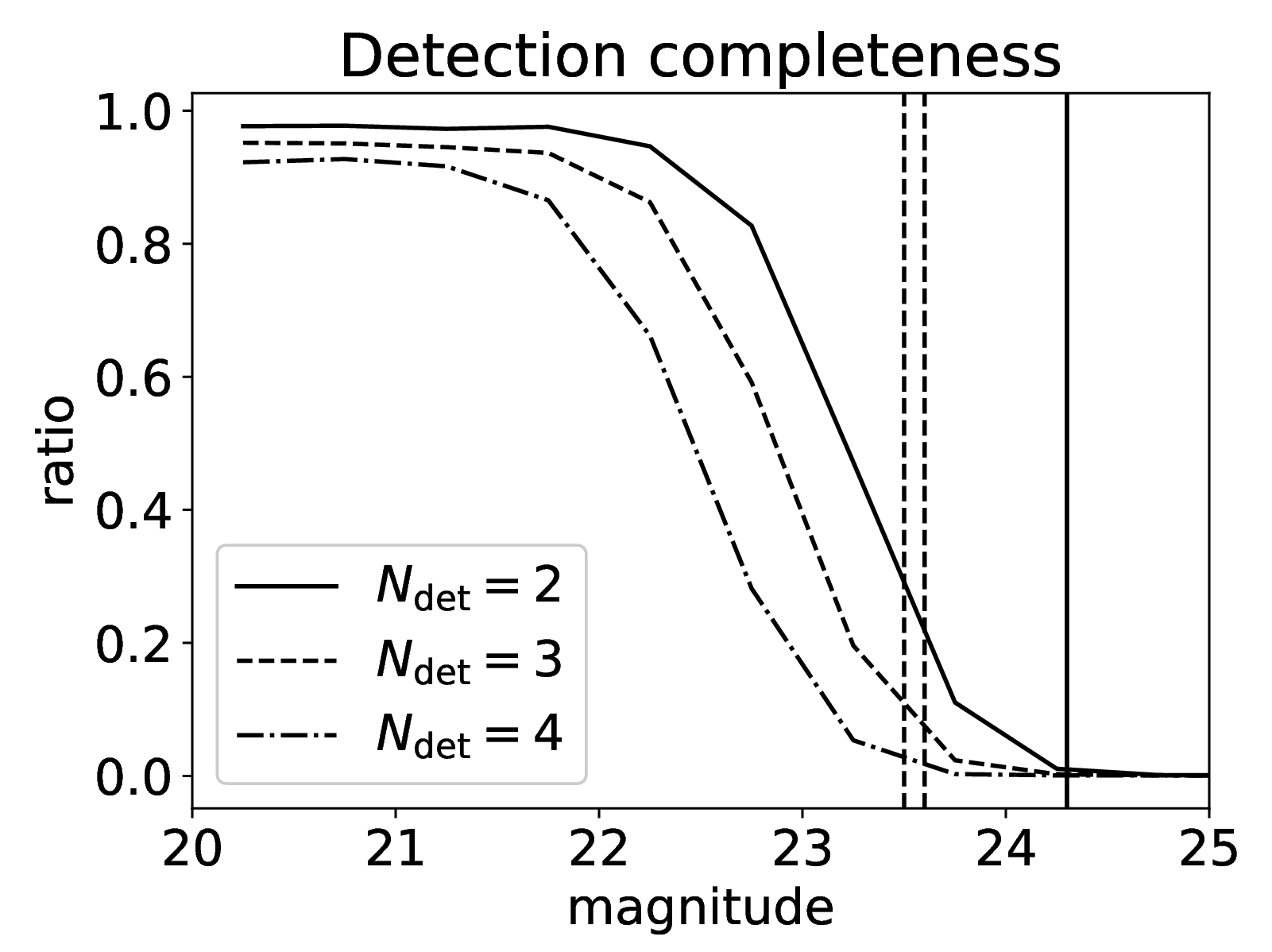} 
 \end{center}
\caption{Detection completeness with different transient candidate definitions. The detection completeness varies according to $N_{\rm det}$. The dashed and solid vertical lines represent the limiting magnitudes for the $z$- and $i$-band differenced images listed in table \ref{tab:qualityfordifferencedimage}, respectively. }\label{fig:completeness}
\end{figure}

Next, we evaluated the detection completeness.
As our transient survey is based on multiple differenced images,
the detection completeness becomes shallower than that in the usual evaluation.
We employed an artificial object test on the differenced images to evaluate the detection completeness.
We compiled an input catalog of artificial sources
and randomly distributed PSF-shaped sources  with various magnitudes over the entire survey field.
Then, we repeated the same detection and filtering procedure for real sources
and counted the number of detected artificial sources.
Finally, we calculated the ratios of the number of recovered sources to  the number of injected sources  as a function of magnitude.
We also investigated the detection completeness of the adopted $N_{\rm det}$.
Figure \ref{fig:completeness} shows the evaluated completeness as a function of magnitude.
Different curves show different thresholds of $N_{\rm det}$.
As expected, a larger number of detections ($N_{\rm det}=4$) gives a shallower 50\% completeness magnitude of $i\sim22.5$ mag,
while a smaller number ($N_{\rm det}=2$) gives deeper 50\% completeness magnitude of $i\sim23.2$ mag.
As the depth of the $z$-band images is shallower than that of the $i$-band,
the limiting magnitude of the $z$-band mostly regulates the completeness curve.

We replaced the measurements fainter than the $3\sigma$ limiting magnitudes with the $3\sigma$ limiting magnitude. The limiting magnitudes were measured in each patch. The replacement imposes minimum limits on the magnitude differences for faint candidates.

\section{Counterpart identification strategy}\label{strategy}

\begin{figure*}
 \begin{center}
   \includegraphics[width=8cm,bb=0 0 461 345]{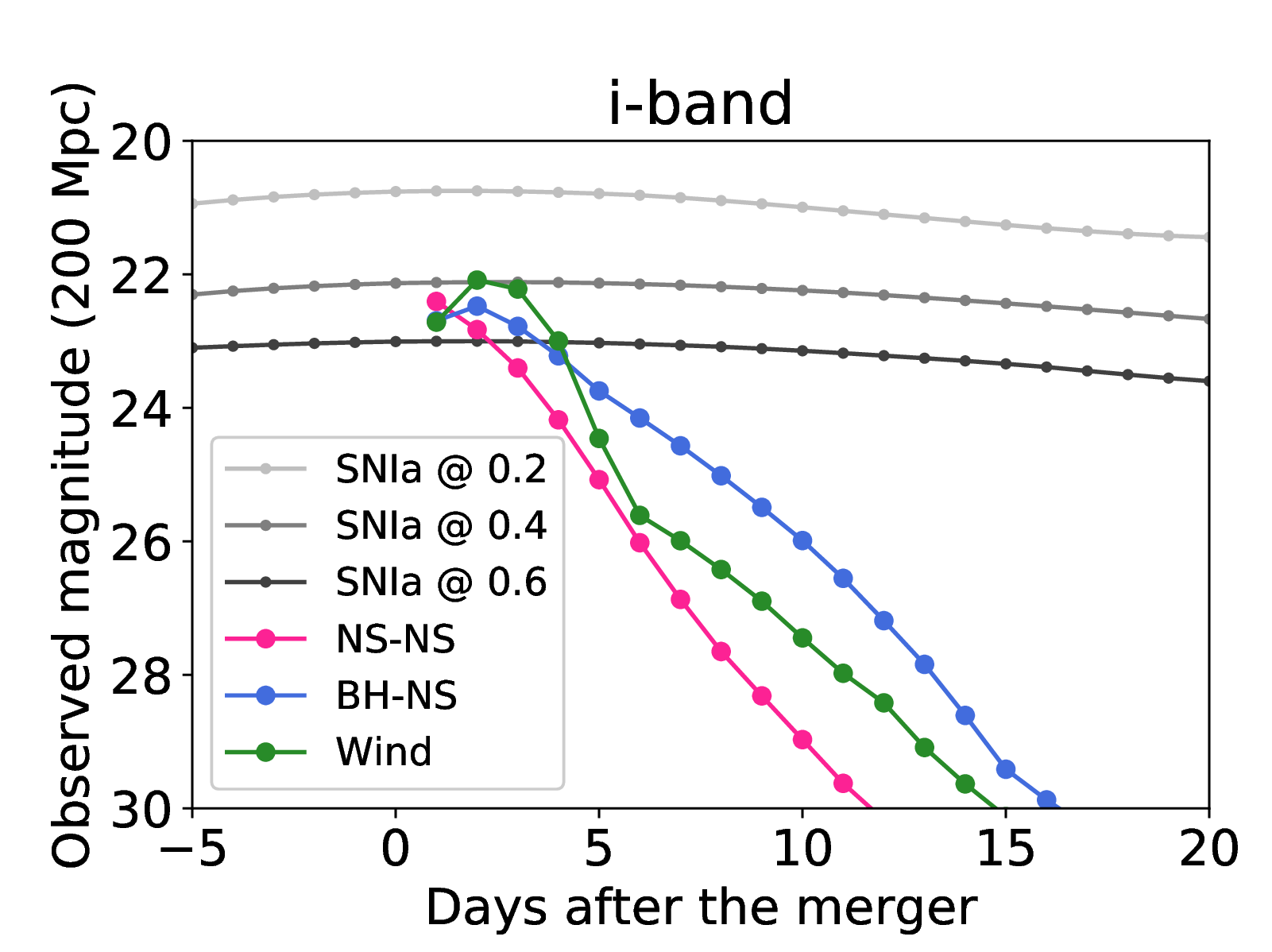}
   \includegraphics[width=8cm,bb=0 0 461 345]{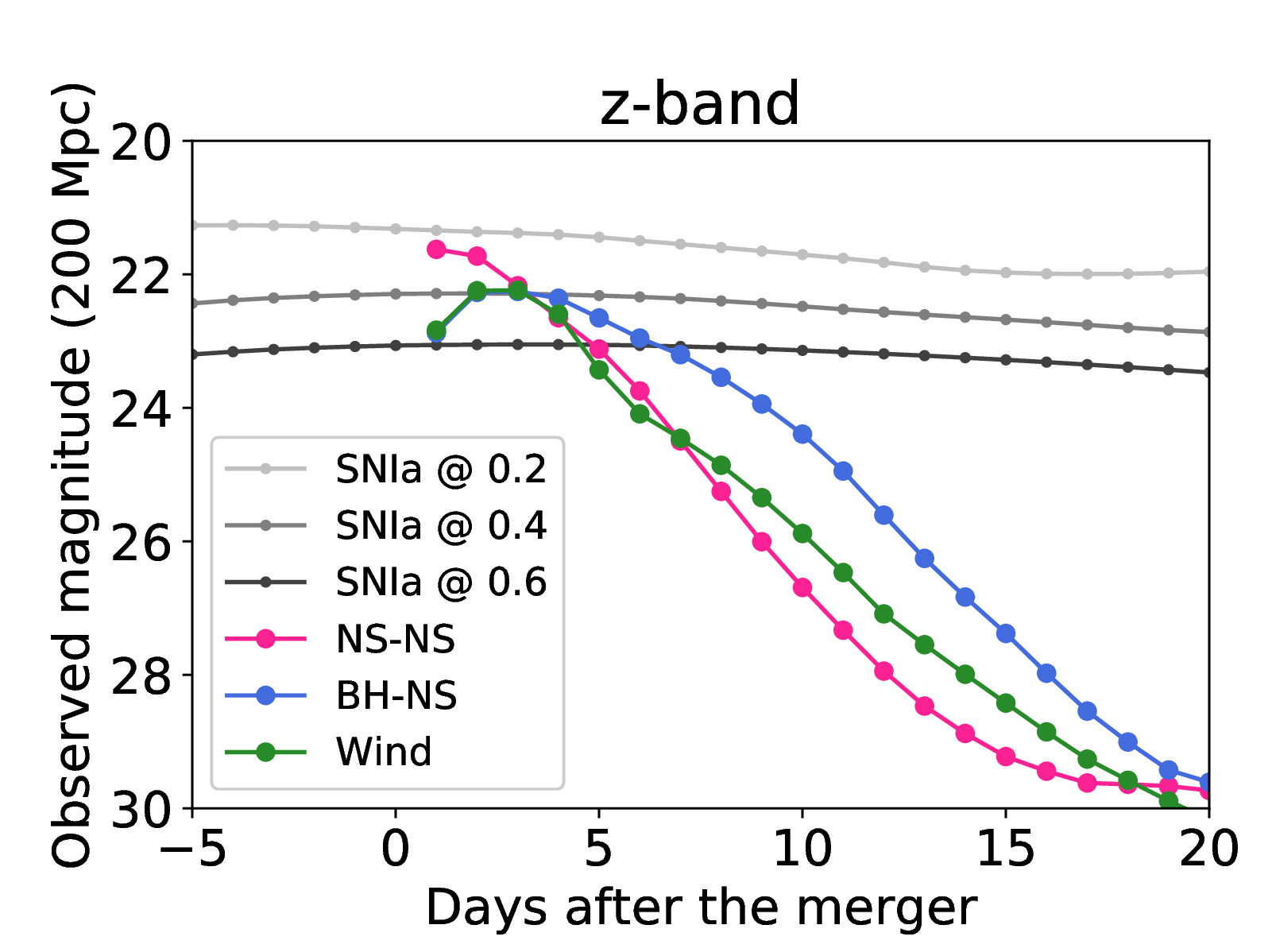}
 \end{center}
\caption{Light curves of kilonova models at 200 Mpc and SNe Ia at different redshifts.
The models for the NS-NS \citep{2013ApJ...775..113T}, BH-NS \citep{2014ApJ...780...31T} and wind \citep{2016AdAst2016E...8T} kilonovae rapidly decrease in  apparent magnitude compared to all the SN Ia models.  For the SN models, t=0 corresponds to the time of maximum light in the B-band. 
}\label{fig:lightcurve}
\end{figure*}

As an optical counterpart to a compact binary coalescence,
we assume kilonova emission, which is powered by
radioactive decays of {\it r}-process nuclei
\citep{1998ApJ...507L..59L,2005astro.ph.10256K,2010MNRAS.406.2650M}.
The timescale of kilonova emission is shorter than that of a typical SN
because of the much lower ejecta mass.
Its spectrum is redder than SNe because of the high opacity of the Lanthanide elements
\citep{2013ApJ...775...18B,2013ApJ...774...25K,2013ApJ...775..113T},
and the flux at longer wavelengths tends to remain relatively high for longer periods than at shorter wavelengths (Figure \ref{fig:lightcurve}),
although
it has been suggested that the mass ejection from the accretion disk formed after the merger,
which is called as a ``wind'' kilonova, can give rise to flux at the shorter wavelengths \citep{2014MNRAS.441.3444M,2015MNRAS.450.1777K}.

\begin{figure*}
 \begin{center}
   \includegraphics[width=8cm,bb=0 0 461 345]{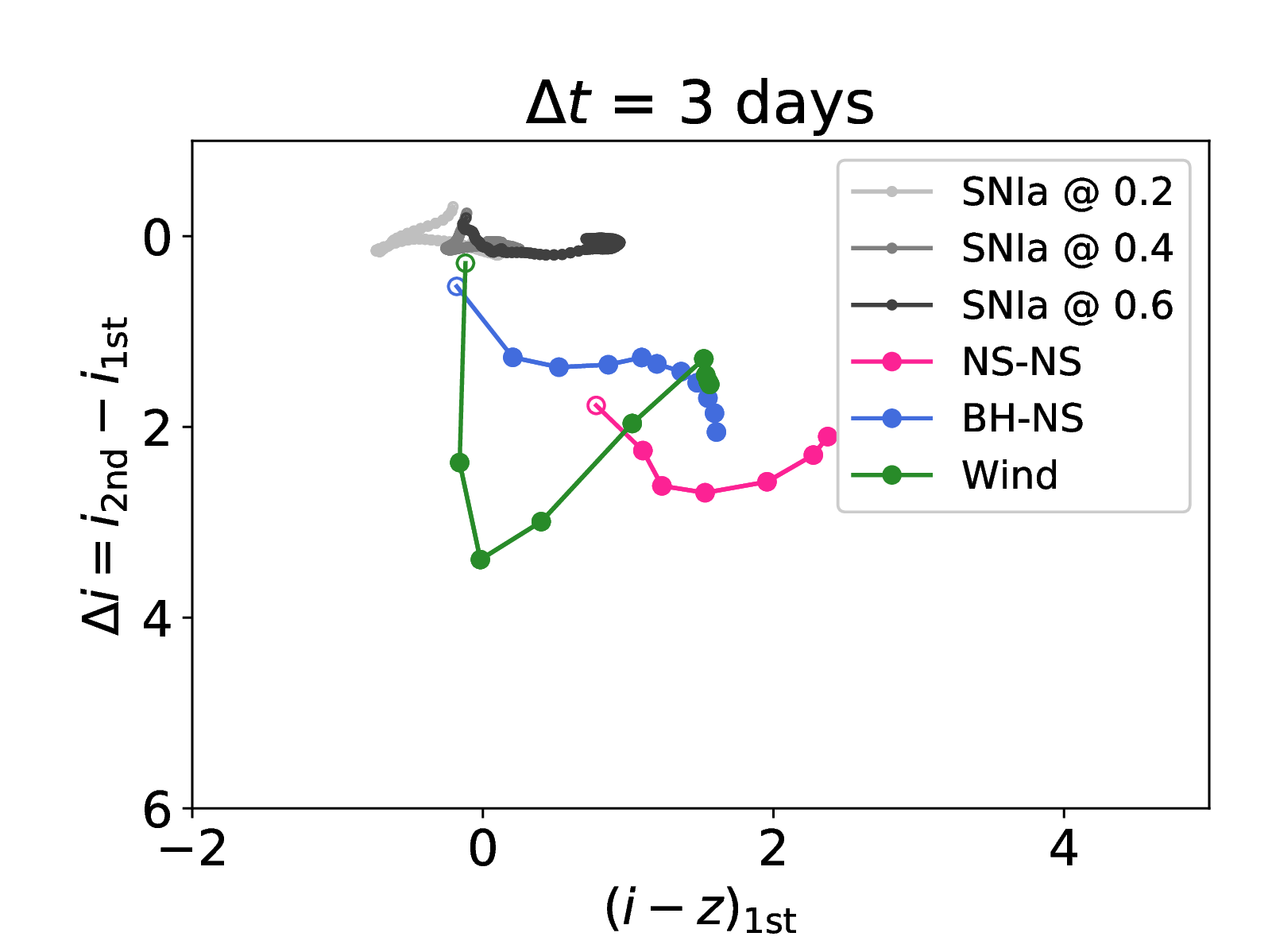}
  \includegraphics[width=8cm,bb=0 0 461 345]{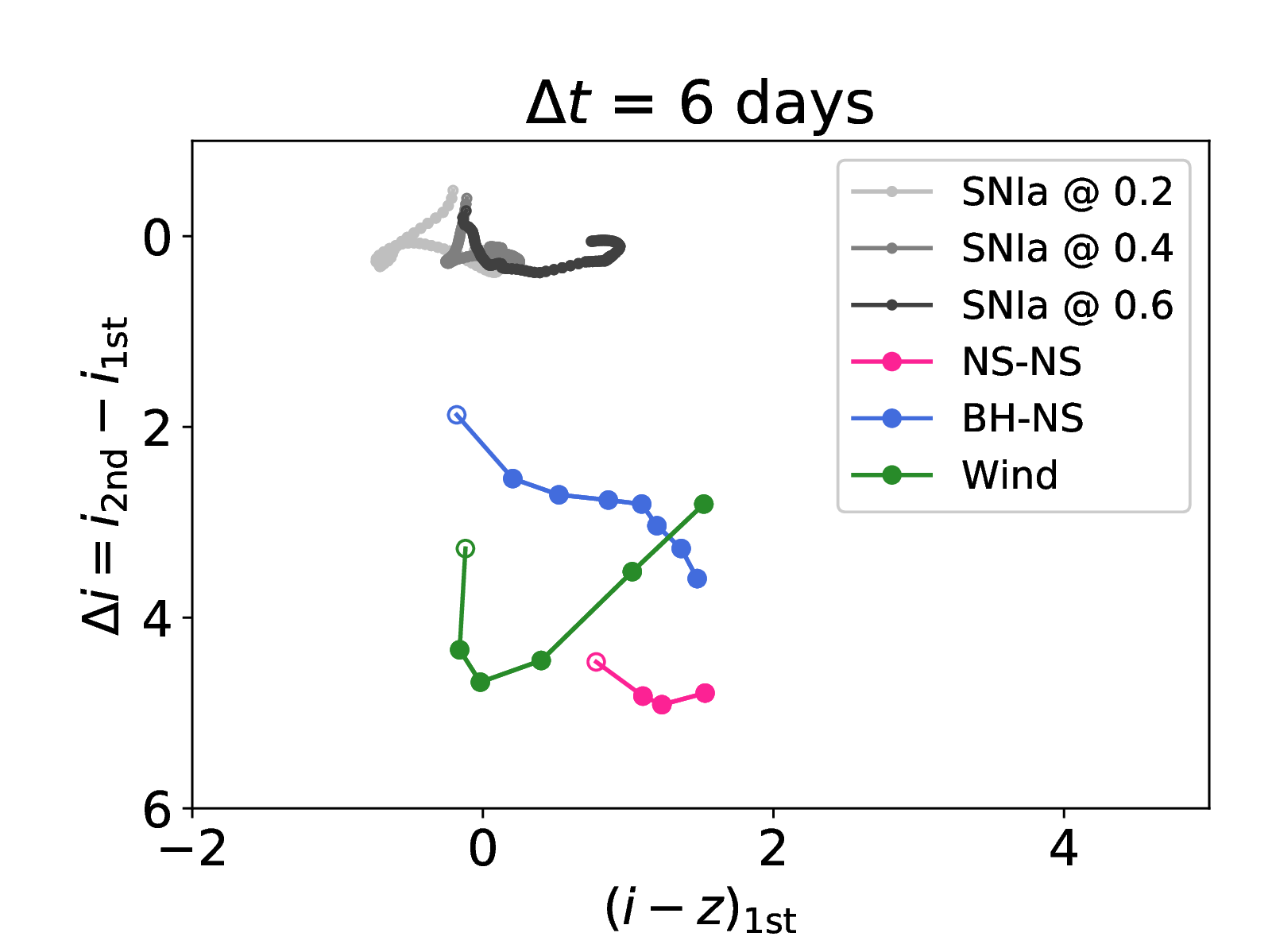}
 \end{center}
\caption{Kilonova and SN models on the $\Delta i$ vs $(i-z)_{\rm 1st}$ plane.
$\Delta i$ vs $(i-z)_{\rm 1st}$ is measured by the difference between the 2nd and 1st epoch ($\Delta t$).
Each dot represents the $\Delta i$ and $(i-z)_{\rm 1st}$ changing its 1st epoch to observe from the first day after the merger with an interval of a day.
The data points for the first day are open circles, while the others are filled circles.
Each kilonova model deviates  from the crowds of SNe.
Even in the case of $\Delta t=3$, the 2nd observation conducted 3 days after the first observation,
the cut of $\Delta i>0.5$ mag provides a good guideline for distinguishing kilonovae from SNe.
}\label{fig:divsizmodel}
\end{figure*}

We investigated the differences in the kilonovae from SN models
in terms of the red color and rapid transient properties \citep[also see][]{2015ApJ...814...25C}.
As Type Ia SNe are the dominant contaminants,
we compare them with models of NS-NS \citep{2013ApJ...775..113T}, BH-NS \citep{2014ApJ...780...31T}, and wind \citep{2016AdAst2016E...8T}
kilonovae.
Type II and Ibc SNe are also possible contaminants; however, it is possible to separate them because
Type II SNe evolve slower \citep[e.g.,][]{2014MNRAS.442..844F}
and Type Ibc SNe are not very different from Type Ia SNe in terms of timescale
\citep[e.g.,][]{2010ApJS..190..418G,2014ApJ...794...23D}.
To quantify the rapid decline, we define the magnitude difference
between the 1st  and the 2nd epochs as $\Delta m = m_{\rm 2nd} - m_{\rm 1st}~(m=i, z)$.
Figure \ref{fig:divsizmodel} shows the models of NS-NS, BH-NS,  and wind kilonovae and Type Ia SNe
on the $\Delta i$ and $(i-z)_{\rm 1st}$ plane, where $(i-z)_{\rm 1st} = i_{\rm 1st} - z_{\rm 1st}$.
The figure clearly indicates that the rapid decline continues
for more than a week with a decline rate of $>0.3$ mag/day,
suggesting that the difference in magnitudes among the two
epochs within a few days can discriminate a kilonovae from SNe.

In addition to the light curve and color of a kilonova,
the association with a nearby galaxy is an important factor
\citep{2016ApJ...820..136G,2016ApJ...829L..15S}
because the maximum distance of a GW detection for a NS-NS merger is
about 200 Mpc based on the designed sensitivity of Advanced LIGO.
The majority of NS-NS binaries merge within 30 kpc of the host galaxies, whereas, 
a non-negligible fraction ($\sim10\%$) should occur well outside of
the galaxies because of the kick velocity during the explosions
\citep[\cite{1999MNRAS.305..763B}, see][for a recent review]{2014ARA&A..52...43B}.

To check the association with galaxies, we used the GLADE catalog v1.2 \citep{2016yCat.7275....0D},
which is an all-sky galaxy catalog with a high completeness for identifying GW sources.
It is a compilation of the following galaxy catalogs: GWGC, 2MPZ, 2MASS XSC, and HyperLEDA.
The estimated completeness of the catalog is complete to 73 Mpc,
with a relatively high completeness (53\%) even at 300 Mpc,
by calculating the integrated luminosity function as a function of distance and comparing these with the local luminosity density.
We adopt a projected physical radius tolerance of 60 kpc in the pairing of a transient candidate and a galaxy.

AGNs are also possible  contaminants, but their position in the host galaxy helps the discrimination,
because they are usually located at the center of galaxies. 
However, as GW events can occur even at the centers of galaxies,
transients at galaxy centers cannot be excluded.
To estimate the AGN contamination ratio, statistical evaluation is required. 
The number of optical transients with a shorter time scale than 10 days for a depth of $i <25$ mag is small,
a few per square degree, and it increases for fainter magnitudes \citep{2008ApJ...676..163M}.
Our survey depth is $1\sim2$ mag shallower than theirs.
With this depth, the expected number of faint AGNs with variability detection is fewer than $1~{\rm deg}^{-2}$

Flares in M-dwarfs are also candidate contaminants for our transient candidates.
A mean flare amplitude of $\Delta u\approx1.5$ mag \citep{2009AJ....138..633K} suggests that smaller amplitudes in the $i$-band are expected
when considering the fact that they are generally brighter in the $u$-band than in the $i$-band because of a typical temperature of about $10^4$ K \citep{2013ApJ...779...18B}.
Actually, in the case of a maximum amplitude of $\Delta u = 5.50$ mag, $\Delta i = 0.77$ mag is observed \citep{2009AJ....138..633K}.
We can measure magnitudes even below the 50\% completeness magnitudes,
because the limiting magnitudes of the pre-differenced image are fainter than those are.
A stellar-like counterpart should be visible in the reference frame if a detected candidate is an M-dwarf flare,
which can be identified by inspecting the presence of a stellar-like counterpart in the reference frame.

Asteroids can also be contaminants.
However, we expect that most  asteroid contaminants are rejected
because we require multiple detections for a candidate by selecting a number of detections, $N_{\rm det}$.
A minimum interval of detections, about 3 h, rejects an asteroid moving faster than about  1 pix / h (0.17 arcsec/pix),
whereas the typical movement speed of main-belt and Jovian Trojan asteroids is about a few tens of arcsec/h.
Even for a distant asteroid at 100 AU, a simple calculation of the movement speed at ($\alpha, \delta)=(44^{\circ}, +20^{\circ}$), which is the farthest region from opposition but overlaps with the HSC survey, is 0.50 arcsec/h.

\section{Results}

We examined our transient candidates discovered by the HSC GW optical counterpart survey
with the idea that a red rapid transient associated with a host galaxy is likely to be a candidate.
Although the GW151226 event has been revealed to be a BH-BH merger which is unlikely to be accompanied by an optical counterpart,
it is important to determine the number of surviving candidates in our criteria for  future follow-ups of GW events.
 We summarize our procedure to remove contaminants as follows,  and give descriptions later:

\begin{description}
\item[1] Detect sources from each differenced image (see Section \ref{selection}).
\item[2] Define candidates from the detected sources by selecting a number of detections $N_{\rm det}\geq2$ for a source.
\item[3] Select sources if a candidate is positively visible in the differenced images: $f_{j}>-\Delta f_{j}$ for $f_{j}$  ($j=i_{\rm 1st}, i_{\rm 2nd}, z_{\rm 1st}, z_{\rm 2nd}$).
\item[4a] Apply the red color cut, $(i-z)_{\rm 1st}$.
\item[4b] Apply the rapid decline cut, $\Delta i$.
\item[4c] Check if a visible stellar-like source exists in the reference frame by requiring ``extendedness=0" (table \ref{tab:numbers}).
\item[5] Check association with the galaxies listed in the GLADE catalog within a projected separation of 60 kpc, and follow {\bf 4abc} (table \ref{tab:numberswithoutGLADE})
\end{description}

\begin{table*}[htdp]
\tbl{Numbers of surviving candidates with different criteria.}{
\begin{tabular}{l|rrrrrr}
\hline
								&	Cuts based on decline \\
Condition							& no cut	&	$\Delta i>0.0$	&$\Delta i>0.5$	&	$\Delta i>1.0$	&	$\Delta i>1.5$	\\
\hline
\hline
All 								& 1744(1729) 	&	829(814)	&	236(221)		&	154(139)		&	 118(109)\\
\hline
fadeout only						& 430(415)	&	218(203)	&	92(77)		&	45(30)		&	24(15)\\
+ $(i-z)_{\rm 1st}>0.0$ 				& 294(279)	&	135(120)	&	55(40)		&	29(14)		&	17(8)\\
+ $(i-z)_{\rm 1st}>0.5$ 				& 185(170)	&	72(57)	&	29(14)		&	16(1)			&	9(0)\\
+ $(i-z)_{\rm 1st}>1.0$ 				& 75(71)		&	22(18)	&	8(4)			&	4(0)			&	2(0)\\
\hline
\end{tabular}
}\label{tab:numbers}
\begin{tabnote}
This table corresponds to the result of the Procedures {\bf 4a} and {\bf b}.
The number of candidates with no stellar-like counterpart in the reference frame is shown in parentheses (Procedure {\bf 4c}). 
\end{tabnote}
\end{table*}%

\begin{table*}[htdp]
\tbl{Numbers of surviving candidates associated with GLADE galaxies.}{
\begin{tabular}{l|rrrrrr}
\hline
								&	Cuts based on decline \\
Condition							& no cut	&	$\Delta i>0.0$	&$\Delta i>0.5$	&	$\Delta i>1.0$	&	$\Delta i>1.5$	\\
\hline
\hline
All 								& 132(130) 	&	67(65)	&	16(14)		&	13(11)		&	  8(7)\\
\hline
fadeout only						& 43(41)		&	23(21)	&	8(6)			&	7(5)			&	3(2)\\
+ $(i-z)_{\rm 1st}>0.0$ 				& 32(30)		&	16(14)	&	6(4)			&	5(3)			&	1(0)\\
+ $(i-z)_{\rm 1st}>0.5$ 				& 21(19)		&	9(7)		&	2(0)			&	2(0)			&	1(0)\\
+ $(i-z)_{\rm 1st}>1.0$ 				& 13(11)		&	5(3)		&	2(0)			&	2(0)			&	1(0)\\
\hline
\end{tabular}
}\label{tab:numberswithoutGLADE}
\begin{tabnote}
This table is the same as table \ref{tab:numbers} but is associated with GLADE galaxies
and  corresponds to the result of Procedure {\bf 5}.
\end{tabnote}
\end{table*}%

\begin{figure}
 \begin{center}
  \includegraphics[width=8cm,bb=0 0 461 364]{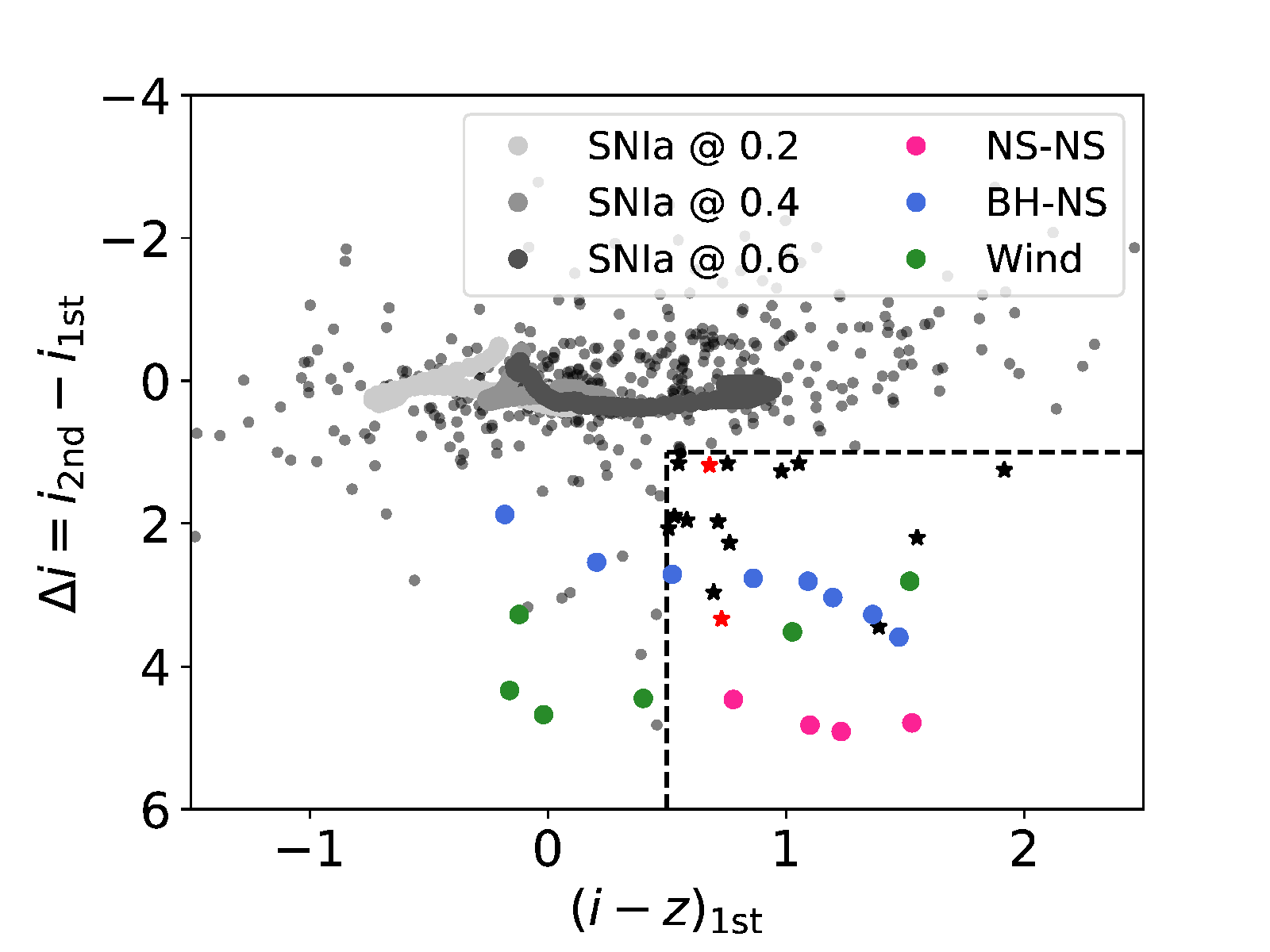} 
 \end{center}
\caption{Transient candidates on the plane of $\Delta i$ vs $(i-z)_{\rm 1st}$ with kilonovae and SNe.
The candidates that survive  Procedure {\bf 3} are shown in this figure.
For the model, 
$\Delta t$ is taken as 6 days, which is the same duration as the HSC observation in this study.
The dashed line indicates the relatively loose cut of ($(i-z)_{\rm 1st}=0.5$ and $\Delta i=1.0$)  as an example (one of the results from applying Procedure {\bf 4b}).
The star symbols indicate the surviving candidates after being cut based on the presence of a stellar counterpart (Procedure {\bf 4c}),
and the star symbols with a red color indicate the candidates located outside the Galactic plane $|b|> 5^{\circ}$.
}\label{fig:dmdcolor}
\end{figure}
There are 1744 candidates that pass our detection criteria requiring twio detections, $N_{\rm det}=2$ (Procedure {\bf 2}).
Among all the detected candidates, we first extract candidates that are fainter in the reference frame (fadeout candidates)
because kilonovae fade out rapidly.  We note that our reference frame are obtained about 42 days later from the LIGO trigger (see Table \ref{tab:stacked}).
The fluxes in the differenced images $f_{j}$  ($j=i_{\rm 1st}, i_{\rm 2nd}, z_{\rm 1st}, z_{\rm 2nd}$)
of the fadeout candidates should be positive or null within $1\sigma$ of uncertainty ($\Delta f_{j}$), that is $f_{j}>-\Delta f_{j}$.
This selection criterion is called ``fadeout only," and 75\% of the candidates are rejected,
resulting in 430 surviving candidates (Procedure {\bf 3}).
A  plot of the transient candidates on the $\Delta i$ and $(i-z)_{\rm 1st}$ plane is presented in Figure \ref{fig:dmdcolor}.

Next, we apply a cut based on the color in the 1st epoch, $(i-z)_{\rm 1st}$  (Procedure {\bf 4a}).
The expected color of a kilonova depends on the date since the explosion.
In tables \ref{tab:numbers} and \ref{tab:numberswithoutGLADE}, we present  three cases with color cuts of $(i-z)_{\rm 1st}=0.0, 0.5,$ and 1.0,
which are likely to be valid for most cases except for the initial few days 
when the coalescences of NS-BH and wind kilonovae fail to meet this red color condition.
According to the color cuts of $(i-z)_{\rm 1st}=0.0, 0.5,$ and 1.0,
the numbers of surviving candidates decrease to 294, 185, and 75, respectively.

The number of candidates is then reduced significantly; however, a certain number of candidates still survive in all cases.
As described in Figure \ref{fig:divsizmodel}, SNe can have a red color as the above condition.
To reduce possible contamination by SNe, we investigate
the number of candidates when we cut the candidates according to the magnitudes between the two epochs, $\Delta i = i_{\rm 2nd} - i_{\rm 1st}$ (decline cut)  (Procedure {\bf 4b}).
The decline cut works well at suppressing the number of SN contaminants
because they do not exhibit such rapid variability,  and it becomes feasible  to visually  inspect 
the number of surviving candidates individually.
For example, the number of candidates satisfying $(i-z)_{\rm 1st}>0.5$ and $\Delta i>1.0$ mag is 16.

We also inspect if a candidate has a stellar-like counterpart in the reference frame (Procedure {\bf 4c}).
The stellar-likeness is judged by a measurement ``extendedness" provided by \emph{hscPipe}.
Although the ``extendedness"  is useful for the star and galaxy separation for a source with a high signal-to-noise ratio,
the measurement of a lower signal-to-noise ratio is not stable.
If at least  one extendedness measurement in any frame is zero,
the candidate is likely to be a star.
Only one candidate survives if we consider this condition, but it looks like a star.

We list the numbers of surviving candidates according to the several conditions in table \ref{tab:numbers}.
A stricter cut of $(i-z)_{\rm 1st}>0.0$, $\Delta i>1.5$ mag, and no stellar-like counterpart results in no candidates. 
In addition, we add another constraint of being associated with galaxies listed in the GLADE catalog,
and the results are presented in table \ref{tab:numberswithoutGLADE} (Procedure {\bf 5}).
This further reduces the number of candidates. 
The parameter space with no candidates becomes larger than $(i-z)_{\rm 1st}>0.5$ and $\Delta i>0.5$ mag.
Even with no color cut, the number of candidates becomes considerably small,
suggesting that the optical counterpart with a bluer color, which can be reproduced by the wind model, could be identified by this analysis.


\begin{figure}
 \begin{center}
   \includegraphics[width=8cm,bb=0 0 461 364]{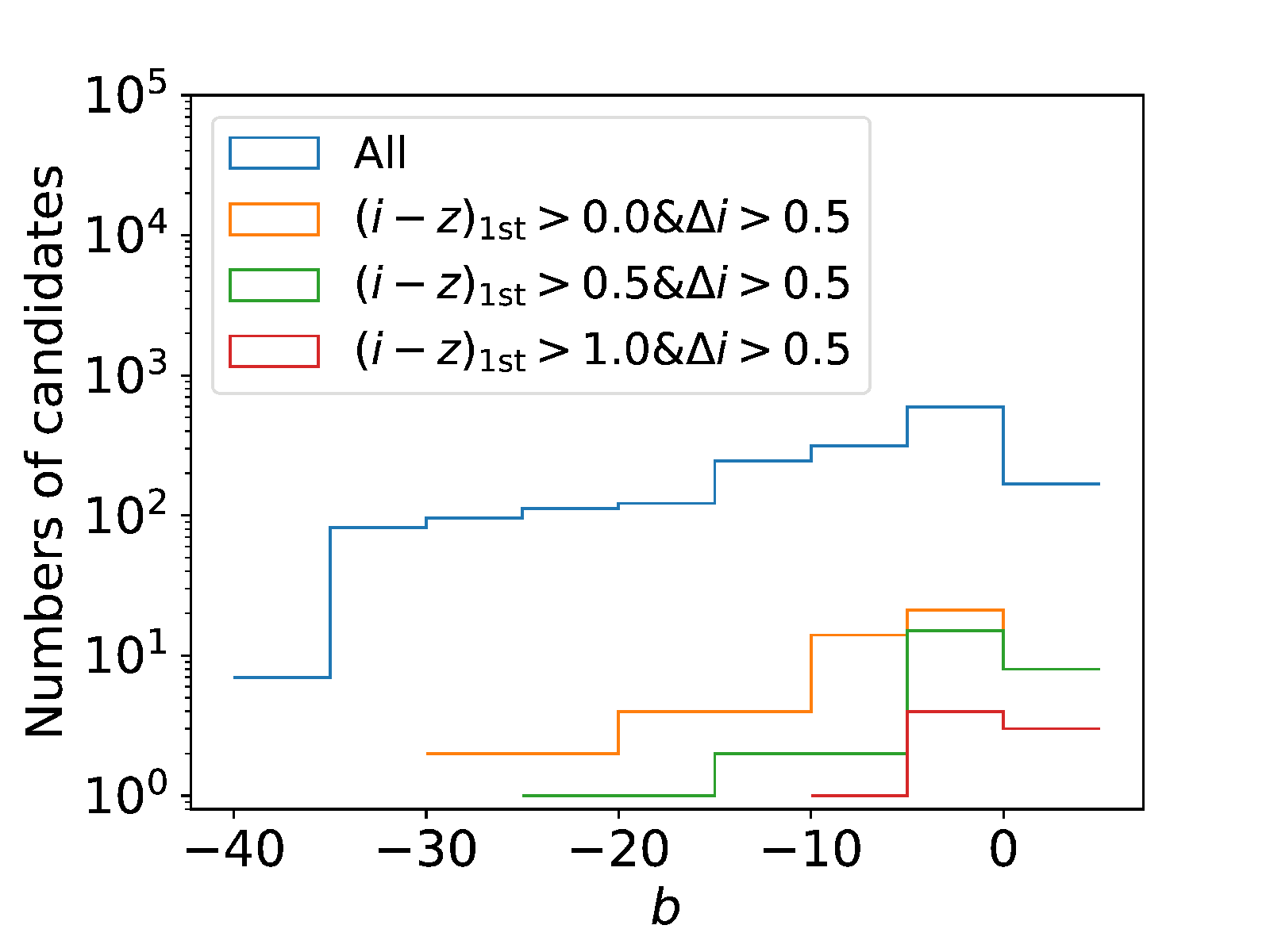} 
 \end{center}
\caption{
Numbers of candidates as a function of Galactic latitude $b$ with different criteria. 
They depend on $b$, and are larger at $b\sim0$, suggesting that the candidates are mostly dominated by Galactic stars}
\label{fig:number_as_b}
\end{figure}
To investigate the properties of the surviving candidates, we plot the number of candidates as a function of Galactic latitude, $b$, as shown in Figure \ref{fig:number_as_b}.
In this figure, we show only the samples in table \ref{tab:numbers}: ``All," ``fadeout only+$(i-z)_{\rm 1st}>0.0$,"  ``fadeout only+$(i-z)_{\rm 1st}>0.5$," and ``fadeout only+$(i-z)_{\rm 1st}>1.0$," and not requiring an association with the GLADE galaxies.
We also require a decline cut of $\Delta i>0.5$ mag with no stellar-like counterpart in the reference frame.
The number of candidates increases toward the Galactic plane, $b\sim0^{\circ}$,
suggesting that our candidates are not extragalactic sources but mostly dominated by Galactic stars.

\begin{figure*}
 \begin{center}

  \includegraphics[width=5cm,bb=0 0 144 144]{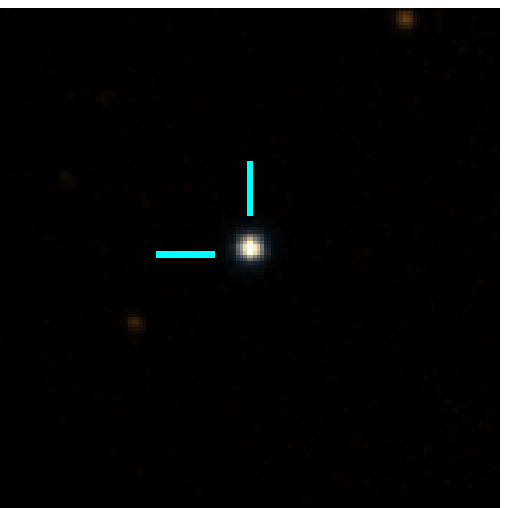}
  \includegraphics[width=5cm,bb=0 0 144 144]{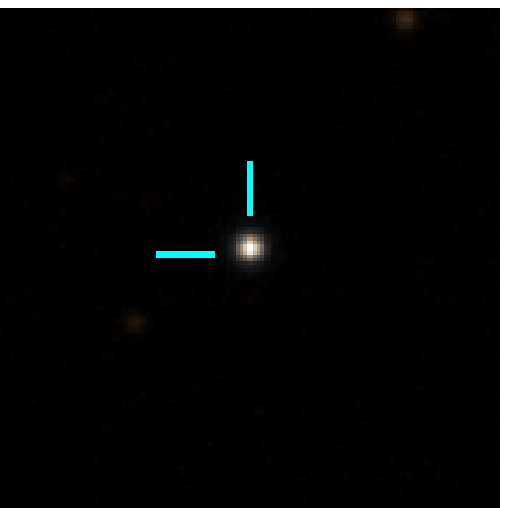}
    \includegraphics[width=5cm,bb=0 0 144 144]{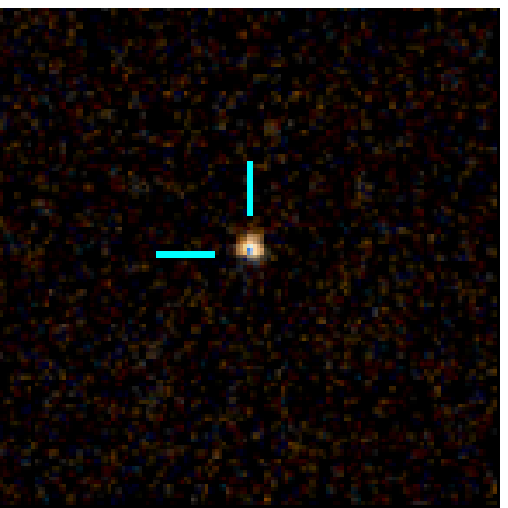}
 \end{center}
\caption{
Candidate with the condition of $(i-z)_{\rm 1st}>0.5$ and $\Delta i>1.0$ (Procedures {\bf 4a} and {\bf b}), but rejected by Procedure {\bf 4c}.
The size of the thumbnails is 20 arcsec $\times$ 20 arcsec.
(left) Pseudo color image of the 1st epoch.
(middle) Pseudo color image of the reference epoch.
(right) Subtracted pseudo color image (left)-(middle).
The color channels of blue, green, and red, are assigned as $i$, $(i+z)/2$, and $z$, respectively.
}\label{fig:thumbnails_of_star}
\end{figure*}

If we adopt a moderately loose cut of $(i-z)_{\rm 1st}>0.5$ and $\Delta i>1.0$ mag,
only two candidates reside offset from the Galactic plane ($|b|>10^{\circ}$)
while the other 16 candidates are within ($|b|<5^{\circ}$).
From visually inspection,
the two are most likely to be stars because the surviving candidates are not associated with the host galaxies but appear to be isolated (Figure \ref{fig:thumbnails_of_star}).
All the other candidates that reside within the Galactic plane also exhibit similar stellar-like features.
In the reference frame, 15 of the 16 candidates have stellar-like counterparts and were rejected by Procedure {\bf 4c}.

\begin{figure*}
 \begin{center}
  \includegraphics[width=5cm,bb=0 0 144 144]{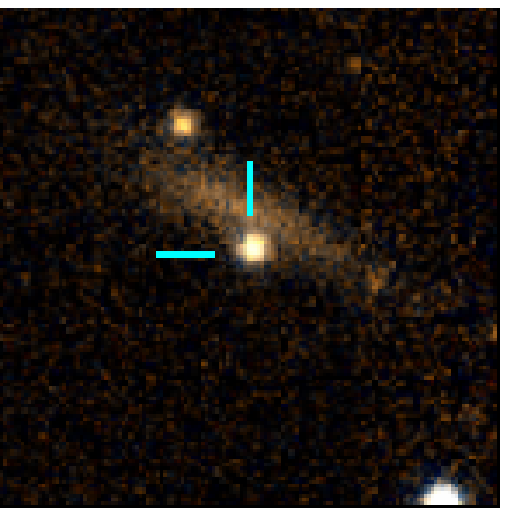}
  \includegraphics[width=5cm,bb=0 0 144 144]{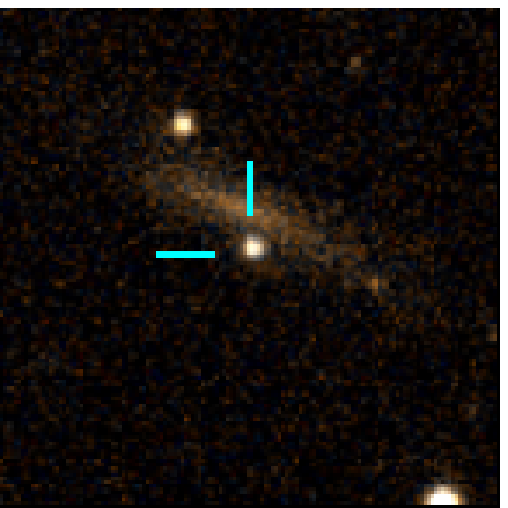}
  \includegraphics[width=5cm,bb=0 0 144 144]{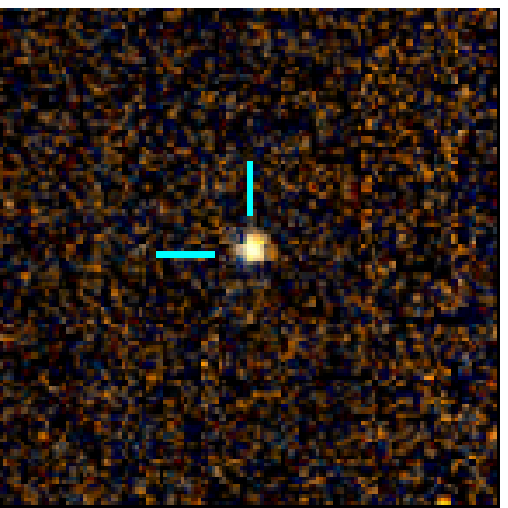}
 \end{center}
\caption{
Same as Figure \ref{fig:thumbnails_of_star} but for 
A Supernova like candidate. It is rejected by the decline cut (Procedure {\bf 4b}).
}\label{fig:thumbnails_of_sn}
\end{figure*}

\begin{figure*}
 \begin{center}
  \includegraphics[width=5cm,bb=0 0 144 144]{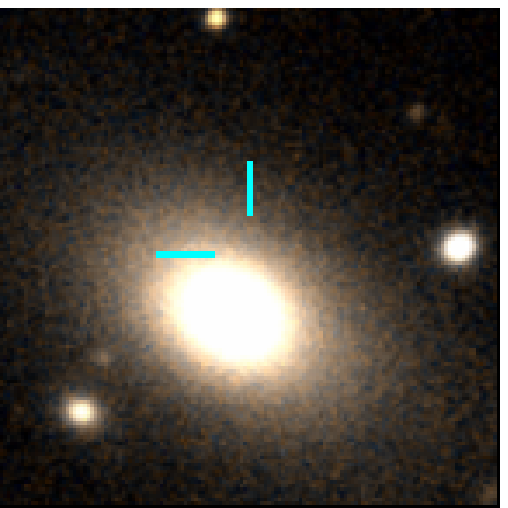}
  \includegraphics[width=5cm,bb=0 0 144 144]{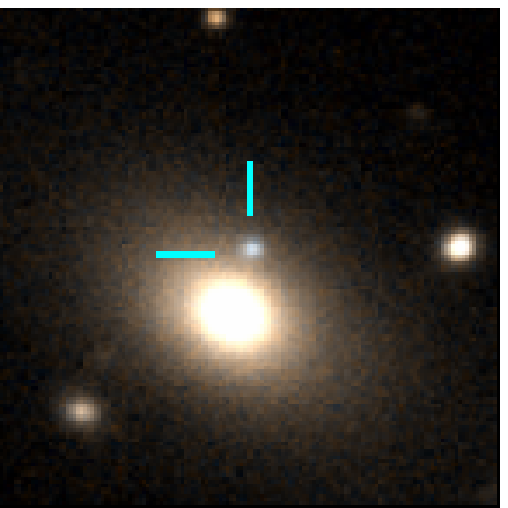}
  \includegraphics[width=5cm,bb=0 0 144 144]{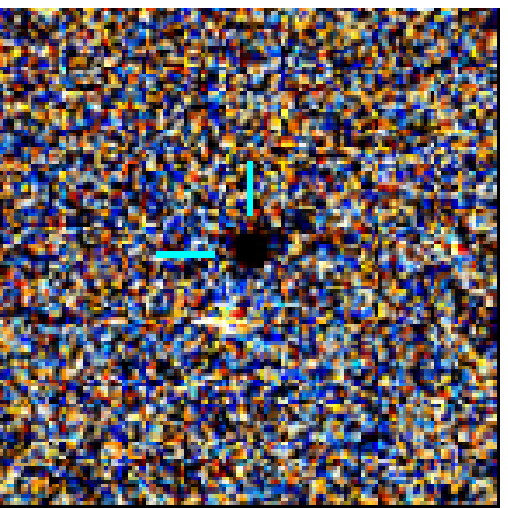}
 \end{center}
\caption{
Same as Figure \ref{fig:thumbnails_of_star} but for 
a candidate associated with a GLADE galaxy. It is rejected by the fadeout cut (Procedure {\bf 3}).
}\label{fig:thumbnails_of_GLADE}
\end{figure*}

We also show thumbnails of a SN-like candidate and a candidate associated with a GLADE galaxy in figures \ref{fig:thumbnails_of_sn} and \ref{fig:thumbnails_of_GLADE}, respectively.
In Figure \ref{fig:thumbnails_of_sn}, the candidate is associated with a clearly visible galaxy, but is offset from the center (left panel). Even in the reference frame, the candidate is still visible (middle panel), suggesting that a light curve with a duration longer than  a month is expected for a SN. The right panel shows the residuals. All other sources surrounding the candidate disappear in the residual image, demonstrating how well the image differencing technique works.
In Figure \ref{fig:thumbnails_of_GLADE}, we show a candidate associated with a GLADE galaxy. As the candidate is not visible in the 1st epoch (even in the 2nd epoch, not shown here) but is visible in the reference frame, it is likely a SN that exploded after the 2nd epoch. This candidate was rejected by Procedure {\bf 3}.

\section{Discussion}

\subsection{Depth of Detection}
The 50\% completeness magnitude of about 23.2 mag only reaches the very early phase of kilonovae at 200 Mpc according to the light curve models (Figure \ref{fig:lightcurve}).
Generally, the depths of detection are different from the measurements.
In particular, the $N_{\rm det}=2$ requirement makes the detection depth one magnitude shallower than the depth of the measurement.
Once a source is identified as ``detected,'' performing forced photometry on each detected source is possible even below the 50\% completeness magnitude.

For example, for the case of a BH-NS merger observed with an interval of 3 days, the magnitude difference will be $\Delta i\sim 1.5-2$ mag.
If we detect a source with $i\sim 23$ mag in the 1st epoch, the 2nd epoch magnitude will be 24.5--25 mag.
As the $3\sigma$ limiting magnitude in the $i$-band is about 25 mag for a 90 s exposure,
we can still measure the magnitude even in the 2nd epoch.
For the case of a NS-NS merger or that with the wind, the situation will deteriorate 
and the magnitude difference will be, $\Delta i\sim 2-3$ mag.
This is too faint to perform  reliable photometric measurements with our wide-field shallow survey strategy.
Even if we fail to obtain photometric measurements of a candidate, we will still be able to derive the lower limit of $\Delta i$.
As our contamination analysis indicates, $\Delta i>1.0$ mag provides a mostly contamination-free sample.
Therefore, the non-detection of a source with a  deep image in the 2nd epoch can also be useful for identifying a kilonova.
The situation is identical to an interval of 6 days.
As the expected difference of magnitudes is larger than the case for a three-day interval, 
the lower limit of $\Delta i$ can be used for identification.

In this study, we investigated the contaminants with an interval of 6 days.
As the interval is a key parameter for discriminating a kilonova from other contaminants, 
the choice of an appropriate interval is important.
 Preliminary studies indicate that an interval of 6 days is acceptable, but further investigations with real data are required for longer intervals.

\subsection{Nature of surviving candidates}
\begin{figure*}
 \begin{center}
   \includegraphics[width=8cm,bb=0 0 461 364]{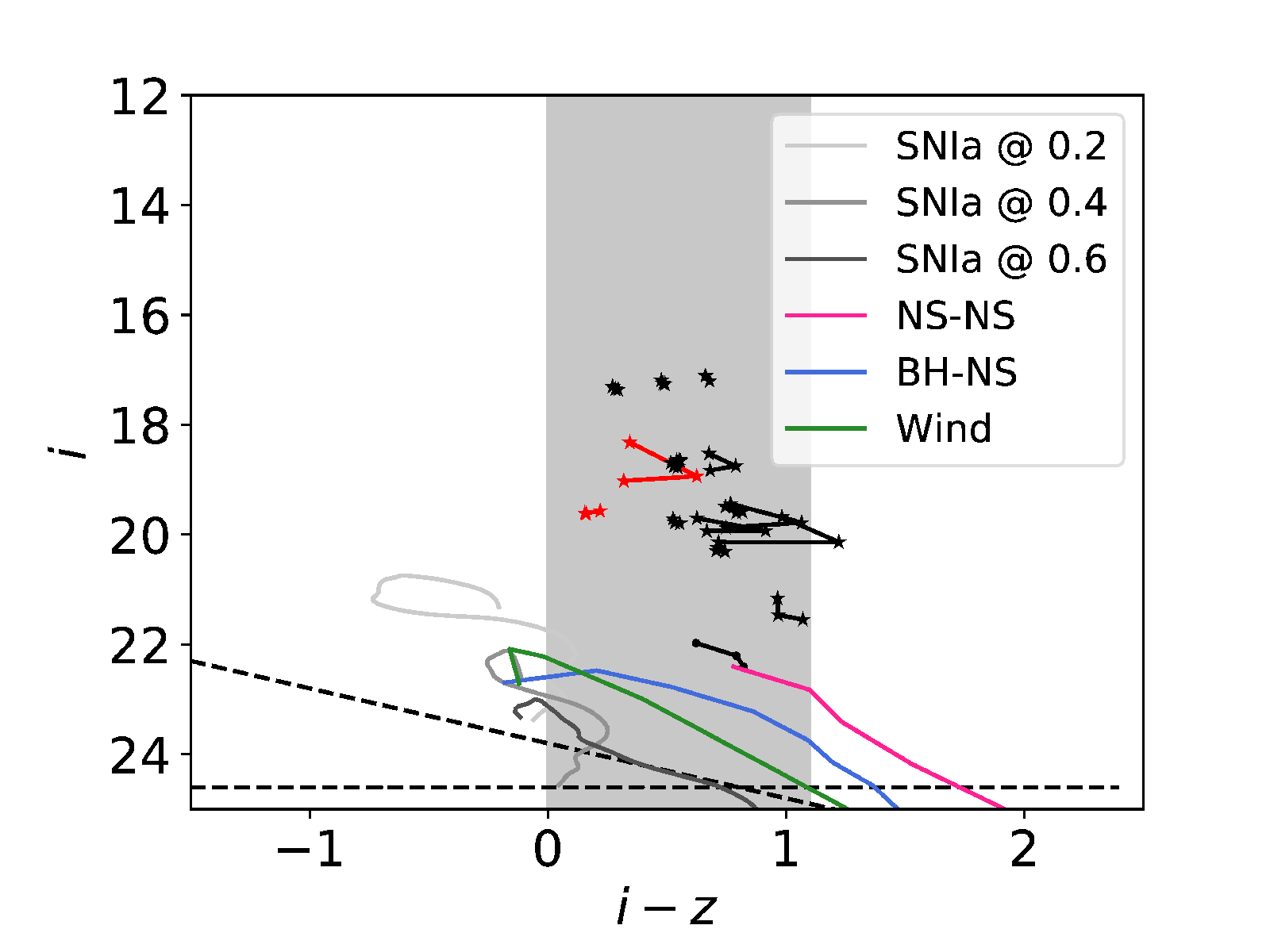} 
   \includegraphics[width=8cm,bb=0 0 461 364]{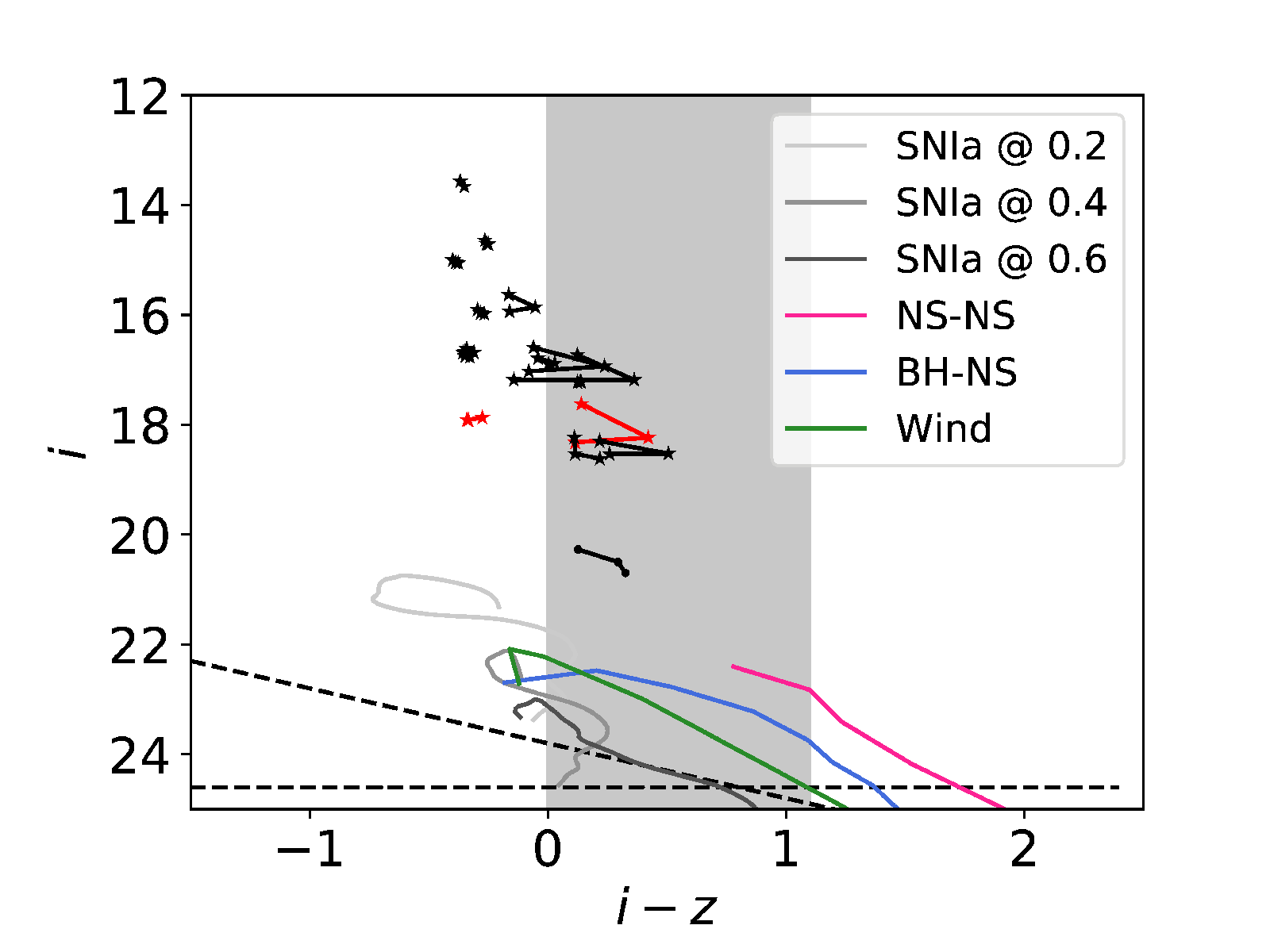} 
 \end{center}
\caption{Color-magnitude diagram for the surviving candidates ($(i-z)_{\rm 1st}>0.5$ and $\Delta i>1.0$; Procedures {\bf 4a,b}).
The left panel shows the color-magnitude diagram without Galactic extinction correction, while the right panel shows that with Galactic extinction correction using a dust map \citep{1998ApJ...500..525S}.
The selection condition is the same as in Figure \ref{fig:dmdcolor}.
The red stars denote candidates residing offset from the Galactic plane $|b|>5^{\circ}$, while the black stars denote the others.
The dots are candidates without a stellar-like counterpart.
The connecting lines represent the paths for each candidate.
All the candidates close to the Galactic plane tend to become brighter and bluer after the Galactic extinction correction.
The shaded region indicates the approximate region for M-dwarfs.
}\label{fig:colormagnitude}
\end{figure*}
Figure \ref{fig:colormagnitude} presents the color-magnitude diagrams of the surviving candidates with measurements by the forced photometry technique on the pre-differenced images.
The forced photometry technique measures the integrated flux at the position where sources are detected on the differenced image.
For comparison, we also show the color-magnitude diagram with the Galactic extinction correction  \citep{1998ApJ...500..525S} in the right panel of figure \ref{fig:colormagnitude}. It shows large differences in both color and magnitude for several candidates induced by the large Galactic extinction.

In Figure \ref{fig:colormagnitude}, there are two types of candidates.
One exhibits no significant difference in color  and  magnitude.
Our large $\Delta m$ selection tends to detect these bright candidates, which have
similar magnitudes in the 2nd and reference images.
Any kind of phenomenon with small fluctuations can explain these, so their nature remains unclear.
The other fades out and changes in color.
The colors of surviving candidates are in the range of $0.0\lesssim i-z \lesssim 1.2$,
which are roughly consistent with those of the M-dwarf stars \citep{2007AJ....134.2398C},
but are likely contaminated by bluer-type stars than M-dwarfs.
M-dwarfs that have variability are known as flare stars (or UV Ceti variables).
Although the amplitude of the flares is larger for bluer  stars,
0.7 mag in the $i$-band is still acceptable for the flares \citep{2009AJ....138..633K}. 
According to \cite{2009AJ....138..633K}, 
the rate of finding a Galactic flare star is on the order of $1/{\rm deg}^2/h$ for $23^{\circ} < |b| < 45^{\circ}$,
assuming the decaying time scale for the flare star is on the order of an hour.
In our 63.5deg$^2$ survey area, it is not surprising that we find a few flare stars.

The reason for the color shift is unclear.
One possible explanation is a pre-cataclysmic variable \citep[pre-CV;][]{2007A&A...474..205T}.
The pre-CV is a detached binary system in the post-common envelope phase.
The system consists of a white dwarf and a late-type star such as an M-dwarf.
The luminosity of a white dwarf is comparable to that of a late-type star  \citep[e.g.][]{1984ApJ...282..612S,2001ApJ...563..971O},
so the color shift can be caused by their eclipse.

Some of those in the right panel of  Figure \ref{fig:colormagnitude} that are bluer than the color range of M-dwarfs can be explained by earlier types  than by an M-dwarf.
However, applying the Galactic extinction value to correct the extinction in Figure \ref{fig:colormagnitude} may cause overestimation, especially for candidates close to the Galactic plane.
A definitive classification of stellar types is not feasible with limited information.
Other possible explanations for these surviving candidates are stars in the red giant branch,
such as RR Lyr, Cepheids, Semi-Regulars, and Miras (see Figure 8 in \citet{2008JPhCS.118a2010E}),
whose absolute magnitudes are around 0.
Assuming the apparent magnitudes of these surviving candidates are $i\sim 15$,
their distances are roughly estimated to be $d = 10^{(15-0)/5+1} {\rm in~pc} = 10{\rm ~kpc}$.
This is slightly larger than the size of our Galaxy.

We investigated the spectral energy distributions (SEDs) of the two surviving candidates residing offset from the Galactic plane
using the PanSTARRS catalog \citep{2016arXiv161205560C}, in order to inspect their nature.
The SEDs suggest that one candidate is consistent with an M-dwarf or M-giant.
The other one that changes its color is not consistent with a simple stellar spectrum, but requires an additional component in the $g$-band.
The $g$-band excess may be caused by a flare or pre-CV signature, or a combination of both.

\subsection{Future prospects}
The reason for no SNe or AGNs appearing in our final candidate is clear:
both the color and decline cuts significantly reduce the contaminants.
Actually, there are candidates associated with a galaxy (possibly a SN) or are in the central part of a galaxy (possibly an AGN)
in the subgroup ``fadeout only"  without color and decline cuts or with a looser color cut of $(i-z)_{\rm 1st}>0.5$ and no decline cuts.
For future follow-up observations,
the conditions $(i-z)_{\rm 1st}>0.5$, $\Delta i>1.0$ mag, and the absence of stellar-like counterparts in the reference frame 
can be used for identifying a  candidate for a kilonova.

The red cut might not be a good criterion, because it may miss wind kilonovae or the early phase of a BH-NS merger as the models predict bluer colors.
Even for such cases, the association with nearby galaxies helps to identify GW sources.
The number of candidates associated with galaxies in the GLADE catalog with no stellar-like counterpart in the reference frame is quite small,
even if we apply  no color cut to the subgroup ``fadeout only." 
However, the GLADE catalog is incomplete for nearby galaxies (53\% within 300 Mpc).
We may miss some of the candidates because of the insufficient completeness.
Improving the completeness of a nearby galaxy catalog is another route to completing the optical counterpart search.

The current large probability map of approximately $1000{\rm ~deg}^2$ is still challenging 
because the capability of the HSC survey is limited to approximately $60~{\rm deg}^2$ per half a night with the current depth of $i\sim 23$ mag.
It is highly desired that the accuracy of the localization is improved  by the participation of other GW interferometers such as Virgo, KAGRA   and LIGO-India.
When the localization accuracy improves, to about 10 deg$^2$, we will be able to probe deeper by spending exposures on a small patch of a possible region.
In contrast, adding additional information is an alternative way in this case:
\citet{2006ApJ...638..354B} suggested that some of the GW events could be associated with short gamma ray bursts (sGRBs).
In addition, there is an expectation that fast radio bursts (FRBs) are associated with kilonovae \citep{2013PASJ...65L..12T}.
Using the association with sGRBs or FRBs will help to improve the localization within the typical error radius of a few arcmin.

Currently,  reference images that have a comparable depth to that of HSC are limited to small areas of the sky.
In order to distribute an immediate notice for further investigation by other facilities, 
it is important to prepare deep sizable reference images prior to  observations.
Differencing with the PanSTARRS $3\pi$ imaging data \citep{2016arXiv161205560C},
whose $i$-band limiting magnitude is about 23.1 mag, may be a good candidate
although it is slightly shallower than the HSC depth. 
Alternatively, performing a very-wide-field imaging survey with HSC on the order of 10000 deg$^2$,
wider than the HSC SSP survey, will increase the chances of achieving rapid identification.

\section{Summary}
Thanks to the largest-in-class aperture size of the telescope,
the HSC observation provides the deepest imaging survey for the optical counterpart search for GW151226.
In half a night, the HSC survey imaged 63.5 deg$^2$ in the $i$- and $z$-band with limiting magnitudes of 24.3--24.6 and 23.5--23.8 mag, respectively.
These limiting magnitudes are apparently the deepest among the optical deep and wide field imagers such as 
DECam \citep{2016ApJ...826L..29C} and PanSTARRS \citep{2016ApJ...827L..40S}.

The limiting magnitude in a pre-differenced image is not a measure of the depth in a transient survey,
because candidates are identified by differenced images.
Our artificial object test gives a 50\% completeness magnitude of approximately 23.2  with $N_{\rm det}=2$, which is shallower than the limiting magnitude.
The theoretical models give estimates of their magnitudes as 22 at a distance of 200 Mpc.
The derived limiting magnitude in a pre-differenced image is still acceptable for measuring the magnitude in the 2nd epoch
although the model predicts a large magnitude difference between observing epochs.
Our HSC survey demonstrates that the large collecting power of HSC is powerful enough to search for  kilonovae.

Assuming a kilonova as an optical counterpart to a GW event,
the color and decline cuts can be used to discriminate a kilonova from other transient candidates.
Even by applying a looser cut, $(i-z)_{\rm 1st}>0.5$ and $\Delta i>1.0$ mag,
the number of transient candidates is reduced to the level of a few percent.
The surviving candidates are identified as stars because they are still visible in the reference frame and are not associated with a galaxy but are isolated.
An efficient way to reduce contaminants is to check the association with a galaxy using the nearby galaxy catalog, GLADE.
Only one candidate is found with the cuts of $(i-z)_{\rm 1st}>0.5$, $\Delta i>1.0$ mag, and no stellar-like counterpart.
The candidate looks similar to a star, which may be due to the failure of the star-and-galaxy separation.
Unfortunately, our observed area overlaps with the Galactic plane, and a large number of flare star contaminants are found.
The large dust extinction prevented us from identifying an extragalactic  kilonova in the Galactic plane.
All the other potential contaminants such as SNe and AGNs could be rejected by our scheme.

When a better probability map is provided by future GW experiments,  with a localization accuracy of about 50deg$^2$,
the EM counterpart search with HSC will become possible by conducting immediate follow-up observations with an interval of 3 days.

\begin{ack}
We are very grateful to all the staff of the Subaru Telescope.
We thank Dr. Hiroyuki Maehara who provided insightful comments on understanding the survived candidates.
We thank Dr. Marica Branchesi for a careful reading of the manuscript.

This work was supported by
the Grant-in-Aid for Scientific Research on Innovative Areas from the MEXT (JP24103003, JP15H00788),
the Grant-in-Aid for Young Scientists (B) from the JSPS (JP25800103, JP26800103),
the Grant-in-Aid for Scientific Research (A) from the JSPS (JP15H02075, 16H02183),
the research grant program of the Toyota Foundation (D11-R-0830),
the natural science grant of the Mitsubishi Foundation, and
the research grant of the Yamada Science Foundation.

The Hyper Suprime-Cam (HSC) collaboration includes the astronomical
communities of Japan and Taiwan, and Princeton University.  The HSC
instrumentation and software were developed by the National
Astronomical Observatory of Japan (NAOJ), the Kavli Institute for the
Physics and Mathematics of the Universe (Kavli IPMU), the University
of Tokyo, the High Energy Accelerator Research Organization (KEK), the
Academia Sinica Institute for Astronomy and Astrophysics in Taiwan
(ASIAA), and Princeton University.  Funding was contributed by the
Ministry of Education, Culture, Sports, Science and Technology (MEXT),
the Japan Society for the Promotion of Science (JSPS), 
Japan Science and Technology Agency (JST),  the Toray Science 
Foundation, NAOJ, Kavli IPMU, KEK, ASIAA,  and Princeton University.

We used software developed for the Large Synoptic Survey Telescope. We thank the LSST Project for making their code available as free software at http://dm.lsstcorp.org \citep{2008arXiv0805.2366I,2010SPIE.7740E..15A}.

Funding for SDSS-III has been provided by the Alfred P. Sloan Foundation, the Participating Institutions, the National Science Foundation, and the U.S. Department of Energy Office of Science. The SDSS-III web site is http://www.sdss3.org/.

SDSS-III is managed by the Astrophysical Research Consortium for the Participating Institutions of the SDSS-III Collaboration including the University of Arizona, the Brazilian Participation Group, Brookhaven National Laboratory, Carnegie Mellon University, University of Florida, the French Participation Group, the German Participation Group, Harvard University, the Instituto de Astrofisica de Canarias, the Michigan State/Notre Dame/JINA Participation Group, Johns Hopkins University, Lawrence Berkeley National Laboratory, Max Planck Institute for Astrophysics, Max Planck Institute for Extraterrestrial Physics, New Mexico State University, New York University, Ohio State University, Pennsylvania State University, University of Portsmouth, Princeton University, the Spanish Participation Group, University of Tokyo, University of Utah, Vanderbilt University, University of Virginia, University of Washington, and Yale University.

The authors wish to recognize and acknowledge the very significant cultural role and reverence that the summit of Maunakea has always had within the indigenous Hawaiian community.
We are most fortunate to have had the opportunity to conduct observations from this sacred mountain.
\end{ack}

\begin{appendix}
\section{Observational limitations}\label{limitations}
The deep and wide-field imaging capability of HSC matches well with the optical follow-up of GWs.
However, there are several limitations to be considered for conducting HSC surveys.
First, HSC is not always available, because the Subaru telescope is not a dedicated telescope for surveys and is accompanied by various instruments and other programs. 
Executing the HSC follow-up is only possible if it is mounted on the Subaru telescope.
In fact, HSC was not available when the notice of GW150914 was issued.
Currently, HSC is equipped on the telescope among gray-to-gray nights through dark nights, roughly two weeks in a month.
A subsystem of HSC, the filter exchanger system, is not operational for the first and last night of each run ($\sim 86\%$ for two weeks is usable).
In addition, requesting the ToO mode, which we use for this GW follow-up,
is only allowed in allotted nights of the Subaru Telescope except for dedicated time exchange nights ($\sim 80\%$ of   Subaru S17A semester  is not allocated for time exchange nights\footnote{https://www.subarutelescope.org/Science/SACM/SAC2016/20161124.ps}).
Immediate follow-ups can be performed if all the  scheduling conditions are satisfied.
Considering those conditions, the duty cycle that can perform the GW follow-up with HSC during the GW observing run is about 30\%.

Even if the ToO observation is scheduled, there are three limitations to be considered for executing observations.
The major limitation is the filter exchange. Exchanging the filter takes 30min.
There are several steps to complete the filter exchange sequence: pointing the telescope to the zenith, closing the mirror covers for safety, 
running the actual exchanging sequence, targeting the field again, and focusing.
Due to this limitation, the observer should minimize exchanging the filters as much as possible at night for better efficiency.
The other limitation is the exposure intervals.
The current exposures interval is about 34 s \citep{2012SPIE.8446E..62U}.
Longer exposures are preferable for higher efficiencies.
The final issue is that HSC does not ensure an imaging quality higher than about 80 deg,
because the instrument rotator cannot follow rapid rotation for higher elevations.
In addition,  elevations that are lower than about 20 deg are not preferable for obtaining better imaging qualities because of the poor natural seeing.

\section{Observation Planning}\label{plan}
Once the pointing definition is completed, the next step involves determining the observation sequence of the pointings.
We constructed  a plan for the observation, taking into account the elevation limitation, $20^{\circ}\lesssim {\rm El} \lesssim 80^{\circ}$.
Here, we describe our idea for the first night as an example.
We began the observation with the $z$-band because the twilight effect is smaller than that of the $i$-band.
As the most probable region passes the meridian during the beginning of the night,
the higher elevation limitation became a problem.
In particular, most of the central and western parts of pointings were above 80 deg for the first half of the night,
so those pointings should be completed as soon as possible from the beginning.
Therefore we initially focused on visiting the pointings of $\alpha\sim 50^{\circ}\rightarrow 45^{\circ}$ indicated with dark circles in the left panel of Figure \ref{fig:pointingmap}.
Once the west part went above 80 deg, we switched to the eastern part ($\alpha\sim 55^{\circ}\rightarrow70^{\circ}$)
and then went back to the central part ($\alpha\sim 55^{\circ}$) and remained at the western part ($\alpha\sim 45^{\circ}\rightarrow 40^{\circ}$).
As the pointings had already passed the meridian, it became mostly straightforward to perform the $i$-band observation sequence.
We started from the central, most probable part toward the western part ($\alpha\sim 55^{\circ}\rightarrow 40^{\circ}$).
Then, we visited the eastern part  ($\alpha\sim 55^{\circ}\rightarrow 70^{\circ}$).
The maximum slewing speed of the Subaru telescope was 0.5 deg / sec while the interval of exposures was 34 sec.
A slewing larger than 17$^{\circ}$ introduces additional overheads.
As the distance from the west end to the west edge of the east part is about 17$^{\circ}$,
we avoided introducing large overheads.

For the first night, we revisited a pointing immediately after the first exposure in order to minimize the slewing overhead.
However, it was difficult to remove asteroids from those data, because the interval of exposures was not sufficient.
To obtain a longer interval, we made exposures on the same pointing, separated as much as possible,
for the remaining 2nd and reference epochs. Once we finished the series of pointings, we revisited it again. 
As a result, the interval reached about the order of an hour.
\end{appendix}

\bibliographystyle{myaasjournal}
\bibliography{GWreferences}

\end{document}